\documentclass[notitlepage,prd,preprintnumbers,showpacs]{revtex4-1}
\usepackage[%
  colorlinks=true,
  urlcolor=blue,
  linkcolor=blue,
  citecolor=blue
]{hyperref}
\usepackage{slashed,color,amsmath,amssymb,mathrsfs}
\usepackage{graphicx}
\usepackage{hyperref}
\usepackage{longtable}
\usepackage{array}
\usepackage{accents}
\usepackage{tensor}

\allowdisplaybreaks

\newcommand{\la}{\langle}
\newcommand{\ra}{\rangle}

\newcommand{\ud}{u^\dag}

\newcommand{\Gam}{\Gamma}
\newcommand{\gamf}{\gamma_5}
\newcommand{\fgamma}{\gamf\gamma}
\newcommand{\psib}{\bar\psi}
\newcommand{\Tb}{\bar{T}}
\newcommand{\Bb}{\bar{B}}
\newcommand*{\ind}{\indices}

\begin{document}
\title{Chiral Lagrangians with decuplet baryons to one loop}
\author{Shao-Zhou Jiang$^{1,2}$}\email{jsz@gxu.edu.cn}
\author{Yan-Rui Liu$^3$}\email{yrliu@sdu.edu.cn}
\author{Hong-Qian Wang$^1$}
\author{Qin-He Yang$^1$}

\affiliation{$^1$Department of Physics and GXU-NAOC Center for Astrophysics and Space Sciences, Guangxi University,
Nanning, Guangxi 530004, People's Republic of China\\
$^2$Guangxi Key Laboratory for the Relativistic Astrophysics, Nanning, Guangxi 530004, People's Republic of China\\
$^3$School of Physics and Key Laboratory of Particle Physics and Particle Irradiation (MOE),
Shandong University, Jinan 250100, People's Republic of China
}

\begin{abstract}
We construct the relativistic chiral Lagrangians with decuplet baryons up to the order $\mathcal{O}(p^4)$ (one loop). For the meson-decuplet-decuplet couplings, there are 1, 13, 55, and 548 terms in the $\mathcal{O}(p^1)$-$\mathcal{O}(p^4)$ order Lagrangians, respectively. For the meson-octet-decuplet Lagrangians, the number of independent terms from $\mathcal{O}(p^1)$ to $\mathcal{O}(p^4)$ are 1, 5, 67, and 611, respectively. For convenience of applications, the $\pi\Delta\Delta$ and $\pi N\Delta$ chiral Lagrangians are picked out. This new form of $\Delta$ Lagrangians is equivalent to the original isovector-isospinor one and we establish relations between these two forms.
\end{abstract}
\maketitle

\section{Introduction}\label{intr}

Chiral perturbation theory (ChPT) is a useful tool to describe low-energy strong interactions of mesons \cite{weinberg,GS1,GS2} and baryons \cite{Gasser:1987rb}. This effective theory is based on the chiral symmetry of QCD and its spontaneous breaking. The interaction terms and various physical quantities in this theory are organized perturbatively by chiral dimension, the order of $p/\Lambda_\chi$ where $p$ represents the typical scale of momentum and $\Lambda_\chi$ is the scale of chiral symmetry breaking. Theoretically, the higher chiral dimension terms are considered, the higher precisions of results would be obtained. At present, the chiral Lagrangians containing the pseudoscalar mesons \cite{GS1,GS2,p61,Bijnens:1999sh,p6p,p6a1,p6a2,tensor1,U3,ourf} and the ground state baryons \cite{Gasser:1987rb,Krause:1990xc,Ecker:1994pi,1998NuPhA.640..199F,Meissner:1998rw,pin4,pib31,pib32,Jiang:2016vax} (both SU(2) and SU(3)) have been already constructed to the sixth and forth order, respectively. Recently, the chiral Lagrangians with $\Delta(1232)$ are also considered up to the forth chiral order \cite{Hemmert:1997ye,Jiang:2017yda}. For the purpose of applications, the current existent chiral Lagrangians are precise enough for theoretical studies on low energy interactions. However, the above investigations missed a kind of particles, the spin-3/2 hyperons.

In reality, a lot of low-energy QCD problems are related to the chiral Lagrangians with decuplet states which are degenerate with the octet baryons in the large $N_c$ limit. Such problems include: the masses of the octet/decuplet baryons and the mass relations between octet/decuplet baryons \cite{Jenkins:1991ts,Lebed:1993yu,Ren:2013dzt,Ren:2013oaa}, the electromagnetic structures of octet and decuplet baryons (magnetic moments, electric quadrupole moments, and electromagnetic form factors) \cite{Jenkins:1992pi,Geng:2009hh,Geng:2009ys,Li:2016ezv}, the meson-octet/decuplet scattering processes \cite{Liu:2007ct,Liu:2010bw}, the transitions from decuplet states to octet states \cite{Butler:1993ht,Li:2017vmq}, lattice studies of baryon properties \cite{Arndt:2003we,Arndt:2003vd,Tiburzi:2004rh}, and so on. Especially, the studies of the transitions between decuplet and octet baryons can shed light on the possible dibaryons \cite{Haidenbauer:2017sws}. The lowest order chiral Lagrangian with decuplet states is obtained easily \cite{Jenkins:1991es}, but we find only fragmentary results for high order terms in the literature (see the references mentioned above). Such Lagrangians are constructed just for special problems one focuses on. A complete and minimal set of Lagrangians with decuplet baryons is still needed. One purpose of this paper is to construct the chiral Lagrangians with the decuplet baryons to one loop (the 4th chiral order) systematically.

In the SU(2) case, we have obtained the chiral Lagrangians with $\Delta$ up to the order $\mathcal{O}(p^4)$ \cite{Jiang:2017yda}, where we use the isovector-isospinor representation \cite{Tang:1996sq} in the isospin space for the Rarita-Schwinger (RS) fields. The application of such Lagrangians is not so convenient in some cases. On the other hand, in the SU(3) case, the decuplet baryons are represented in the flavour space as a totally symmetric tensor $T_{abc}$. Since the $\Delta$ baryons are members of the decuplet representation, the Lagrangians with $\Delta$ can also be expressed with the symmetric tensor. However, it is apparently not straightforward to make a relation between these two formalisms. Another purpose of the present study is to give new chiral Lagrangians with $\Delta$ in the form of $T_{abc}$ ($a,b,c=1,2$) and establish the relations to the former formalism.

This paper is organized as follows. In Sec. \ref{rev0}, we review the building blocks for the construction of the chiral Lagrangians with the mesons, the external sources, and a part of building blocks with baryon fields. In Sec. \ref{stru}, we present the structures of the chiral Lagrangians and give a full  building blocks with baryon fields. In Sec. \ref{const}, the properties of the building blocks, the linear relations of invariant monomials, and the relations between the original chiral Lagrangians with $\Delta$ and the new forms are given. In Sec. \ref{results}, we list our results and present some discussions. Section \ref{summ} is a short summary.

\section{Building blocks in constructing chiral Lagrangians}\label{rev0}
Generally speaking, the constructed Lagrangians in ChPT involve the pseudoscalar mesons, the external sources, the decuplet baryons, and the octet baryons. In this section, we present appropriate building blocks in constructing the chiral Lagrangians. More detailed discussions about them can be found in Refs. \cite{GS1,GS2,Gasser:1987rb,p61,Bijnens:1999sh,p6a2,tensor1,ourf,pin4,pib31,Jiang:2016vax,Hemmert:1997ye,Jiang:2017yda}. For the spin-3/2 baryon states, we consider both $SU(3)$ and $SU(2)$ cases. For convenience, we simply call the form of chiral Lagrangians with $\Delta$ in Ref. \cite{Jiang:2017yda} ``original'' and that in this paper ``new''. Needless to say, the new form $SU(2)$ Lagrangians are just selected terms of the $SU(3)$ Lagrangians with decuplet baryons. Hence, in the following parts, we treat them in the same way.

\subsection{Building blocks of the mesons and the external sources}\label{rev}

The QCD Lagrangian $\mathscr{L}$ can be written as
\begin{eqnarray}
\mathscr{L}=\mathscr{L}^0_{\mathrm{QCD}}+\bar{q}(\slashed{v}+\slashed{a}\gamma_5-s+ip\gamma_5)q~,
\end{eqnarray}
where $\mathscr{L}^0_{\mathrm{QCD}}$ is the original QCD Lagrangian and $q$ denotes the quark field. We use $s$, $p$, $v^\mu$, and $a^\mu$ to denote scalar, pseudoscalar, vector, and axial-vector external sources, respectively. Conventionally, the tensor source and the $\theta$ term are ignored. As usual, we consider that only $a^\mu$ is traceless in the two-flavour case, but both $a^\mu$ and $v^\mu$ are traceless in the three-flavour case.

In ChPT, the pseudoscalar mesons (Goldstone bosons) come from the spontaneous breaking of the global symmetry $SU(N_f)_L\times SU(N_f)_R$ into $SU(N_f)_V$. The resulting meson fields are collected in $u$ and it transforms as
\begin{align}
u\to g_L uh^\dag=h ug_R^\dag
\end{align}
under the chiral rotation, where $g_L$ and $g_R$ represent elements in $SU(N_f)_L$ and $SU(N_f)_R$, respectively, and $h$ is a compensator field which is a function of the pion fields.

To construct the chirally invariant Lagrangians involving only meson fields and external sources, the building blocks are usually chosen as
\begin{align}\label{df}
\begin{split}
u^\mu&=i\{u^\dag(\partial^\mu-ir^\mu)u-u(\partial^\mu-il^\mu)u^\dag\},\\
\chi_\pm&=u^\dag\chi u^\dag\pm u\chi^\dag u,\\
h^{\mu\nu}&=\nabla^\mu u^\nu+\nabla^\nu u^\mu,\\
f_+^{\mu\nu}&=u F_L^{\mu\nu} u^\dag+ u^\dag F_R^{\mu\nu} u,\\
f_-^{\mu\nu}&=u F_L^{\mu\nu} u^\dag- u^\dag F_R^{\mu\nu} u=-\nabla^\mu u^\nu+\nabla^\nu u^\mu,
\end{split}
\end{align}
where $r^\mu=v^\mu+a^\mu$, $l^\mu=v^\mu-a^\mu$, $\chi=2B_0(s+ip)$, $F_R^{\mu\nu}=\partial^{\mu}r^{\nu}-\partial^{\nu}r^{\mu}-i[r^{\mu},r^{\nu}]$, $F_L^{\mu\nu}=\partial^{\mu}l^{\nu}-\partial^{\nu}l^{\mu}-i[l^{\mu},l^{\nu}]$, and $B_0$ is a constant related to the quark condensate. The form of these building blocks, however, is not very useful in the construction of chiral Lagrangians with decuplet baryons. For convenience, we write the flavour indices of these building blocks (or any other matrices in the flavour space) explicitly,
\begin{align}
X={X_a}^b+X_s I,\qquad X_s=\frac{1}{N_f}\la X\ra,\label{dX}
\end{align}
where $X$ denotes any building block in Eq. \eqref{df} (or any matrix in the flavour space), ${X_a}^b$ ($X_s$) is the traceless (traceable) part of $X$, $I$ is the $N_f\times N_f$ identity matrix in the $N_f$-flavour space, and $\la\cdots\ra$ means the trace in the flavour space. We use $a$ and $b$ ($a,b=1,2,3$) to denote the row index and column index of the matrix $X$, respectively. In the following, we will treat the row index (or the first index) of ${X_a}^b$ as the subscript and the column index (or the second index) as the superscript. According to these notations, we have $u^\mu_s=f_{-,s}^{\mu\nu}=h^{\mu\nu}_s=0$ in the two-flavour case and an additional relation $f_{+,s}^{\mu\nu}=0$ in the three-flavour case. The chiral transformations ($R$) for these building blocks are
\begin{align}
\begin{split}
{X_a}^b&\xrightarrow{R} {X'_a}^b={h_a}^{a'}X\ind{^\prime_{a'}^{b'}}h\ind{^\dag_{b'}^b},\\
X_s&\xrightarrow{R}X'_s=X_s.
\end{split}\label{crm}
\end{align}
Here ${h_a}^{a'}$ does not need to be traceless as the definition of ${X_a}^b$ in Eq. \eqref{dX}. The row index of ${X_a}^b$ is related to the $h$ field, but the column index is related to the $h^\dag$ field.

The covariant derivative $\nabla^{\mu}$ acting on the building blocks in Eq. (\ref{crm}) are
\begin{eqnarray}\label{cd}
\begin{split}
\nabla^{\mu}{X_a}^b&=\partial^\mu {X_a}^b+{\Gamma_a}^{c,\mu}{X_c}^b-{X_a}^c{\Gamma_c}^{b,\mu},\\
\nabla^{\mu}X_s&=\partial^\mu X_s,\\
\Gamma^{\mu}&=\frac{1}{2}\{\ud(\partial^\mu-ir^\mu)u+u(\partial^\mu-il^{\mu})\ud\}.
\end{split}
\end{eqnarray}
In constructing the Lagrangian, the following two relations will be useful
\begin{align}
[\nabla^\mu,\nabla^\nu]{X_a}^b&={\Gamma_a}^{c,\mu\nu}{X_c}^b-{X_a}^c{\Gamma_c}^{b,\mu\nu},\label{Gam0}\\
[\nabla^\mu,\nabla^\nu]X_s&=0,\label{Gam00}\\
\Gamma^{\mu\nu}&=\nabla^{\mu}\Gam^{\nu}-\nabla^{\nu}\Gam^{\mu}-[\Gam^{\mu},\Gam^{\nu}]
=\frac{1}{4}[u^\mu,u^\nu]-\frac{i}{2}f_{+}^{\mu\nu}.\label{Gam1}
\end{align}

\subsection{Building blocks of baryons}\label{rev1}
Besides the meson fields and external fields, we also need baryons belonging to $SU(3)$ 8 and 10 representations. The octet baryons are represented by a matrix ${B_a}^b$,
\begin{align}
{B_a}^b=\begin{pmatrix}
\frac{\Sigma^0}{\sqrt{2}}+\frac{\Lambda}{\sqrt{6}} & \Sigma^+ & p\\
\Sigma^- & -\frac{\Sigma^0}{\sqrt{2}}+\frac{\Lambda}{\sqrt{6}} & n\\
\Xi^- & \Xi^0 & -\frac{2\Lambda}{\sqrt{6}}
\end{pmatrix}.
\end{align}
In the two-flavour case, it is reduced to the nucleon doublet,
\begin{eqnarray}
\psi_a=\begin{pmatrix}
p\\n
\end{pmatrix}.
\end{eqnarray}
One may also use the symbol ${B_a}^3$ ($a=1,2$) to denote this nucleon doublet. For the decuplet baryons, they are denoted by a totally symmetrical tensor $T_{abc}$ with
\begin{eqnarray}
&T_{111}=\Delta^{++},\quad T_{112}=\frac{\Delta^+}{\sqrt3},\quad T_{122}=\frac{\Delta^0}{\sqrt3}, \quad T_{222}=\Delta^-,&\nonumber\\
&T_{113}=\frac{\Sigma^{*+}}{\sqrt3},\quad T_{123}=\frac{\Sigma^{*0}}{\sqrt6},\quad T_{223}=\frac{\Sigma^{*-}}{\sqrt3},&\nonumber\\
&T_{133}=\frac{\Xi^{*0}}{\sqrt3},\quad T_{233}=\frac{\Xi^{*-}}{\sqrt3}, \quad T_{333}=\Omega^{-}.&\nonumber
\end{eqnarray}
In the $SU(2)$ case, only the first four fields are needed.

The chiral transformations for these baryon fields are
\begin{align}
\begin{split}
{B_a}^b&\xrightarrow{R} B\ind{^\prime_a^b}={h_a}^{a'}B\ind{_{a'}^{b'}}h\ind{^\dag_{b'}^b},\\
\psi_a&\xrightarrow{R}\psi'_a={h_a}^{b}\psi_b,\\
T_{abc}&\xrightarrow{R}T'_{abc}={h_a}^{a'}{h_b}^{b'}{h_c}^{c'}T_{a'b'c'},\\
{\Bb_a}^b&\xrightarrow{R} \Bb\ind{^\prime_a^b}={h_a}^{a'}\Bb\ind{_{a'}^{b'}}h\ind{^\dag_{b'}^b},\\
\psib^a&\xrightarrow{R}\psi^{\prime a}=\psib^bh\ind{^\dag_b^{a}},\\
\Tb^{abc}&\xrightarrow{R}\Tb^{\prime abc}=\Tb^{a'b'c'}h\ind{^\dag_{a'}^{a}}h\ind{^\dag_{b'}^{b}}h\ind{^\dag_{c'}^{c}}.\\
\end{split}\label{crmB}
\end{align}
From the transformations, the indices of $\psi_a$ and $T_{abc}$ ($\psib^a$ and $\Tb^{abc}$) can be treated as row (column) indices and those of ${B_a}^b$ and ${\Bb_a}^b$ are self-evident. From Eqs. \eqref{crm} and \eqref{crmB}, we can see that if a term is chirally invariant, all the row indices must be contracted with the column indices and vice versa. This is the reason why we write the row and column indices explicitly.

The covariant derivative $D^\mu$ acting on the baryon fields are \cite{1998NuPhA.640..199F,pib31,Jenkins:1992pi}
\begin{align}
\begin{split}
D^\mu\psi_a&=\partial^\mu\psi_a+{\Gamma_a}^{b\mu}\psi_b,\\
D^{\mu}{B_a}^b&=\partial^\mu {B_a}^b+{\Gamma_a}^{c,\mu}{B_c}^b-{B_a}^c{\Gamma_c}^{b,\mu},\\
D^\mu T_{abc}&=\partial^\mu T_{abc}+{\Gamma_a}^{d\mu}T_{dbc}+{\Gamma_b}^{d\mu}T_{adc}+{\Gamma_c}^{d\mu}T_{abd}.
\end{split}
\end{align}
It seems that, in the three (two)-flavour case, we can choose $T_{abc}^\mu$, $\Tb^{abc,\mu}$, ${B_a}^b$, ${\Bb_a}^b$ ($T_{abc}^\mu$, $\Tb^{abc,\mu}$, $\psi_a$, $\psib^a$), and their covariant derivatives as building blocks. But it is a bit more complex for the spin-$3/2$ RS fields, we will discuss this issue in the next section.

\section{Structures of chiral Lagrangians with decuplet baryons}\label{stru}

A similar discussion in this section has been presented in Ref. \cite{Jiang:2017yda}. Here we only list the necessary ingredients for the Lagrangian construction. More details can be found in Refs. \cite{Rarita:1941mf,Moldauer:1956zz,1958NCim....9S.416F,Aurilia:1969bg,VanNieuwenhuizen:1981ae,Williams:1985zz,Benmerrouche:1989uc,Pascalutsa:1994tp,Haberzettl:1998rw,Pilling:2004cu,2006PAN....69..541K,Tang:1996sq,Hemmert:1996xg,Hemmert:1997wz,Hemmert:1997ye,Hacker:2005fh,Krebs:2009bf,Scherer:2012xha,pin4,pib31,pib32,Jiang:2016vax}.

In this paper, we adopt the vector-spinor representation $\Psi^\mu$ ($\mu=0,1,2,3$) \cite{Rarita:1941mf} for the spin-3/2 fields. The general Lagrangian for a free RS field with mass $m$ reads \cite{Moldauer:1956zz}
\begin{eqnarray}\label{rsl}
\mathscr{L}_{\mathrm{f}}&=&\bar{\Psi}_{\mu}\Lambda_A^{\mu\nu}\Psi_\nu,\\
\Lambda_A^{\mu\nu}&=&-\big[(i\slashed{\partial}-m)g^{\mu\nu}+iA(\gamma^\mu\partial^\nu+\gamma^\nu\partial^\mu)\nonumber\\
&&+\frac{i}{2}(3A^2+2A+1)\gamma^\mu\slashed{\partial}\gamma^\nu\nonumber\\
&&+m(3A^2+3A+1)\gamma^\mu\gamma^\nu\big],\nonumber
\end{eqnarray}
where $A\neq-1/2$ is an arbitrary real number. From this Lagrangian, one derives the equation of motion (EOM) and two subsidiary conditions
\begin{align}
&(i\slashed{\partial}-m)\Psi_\mu=0,\label{eompsif}\\
&\gamma^\mu\Psi_\mu=0,\label{eoms1}\\
&\partial^\mu\Psi_\mu=0.\label{eoms2}
\end{align}
The two unphysical spin-$\frac{1}{2}$ degrees of freedom in the vector-spinor representation can be eliminated with these two subsidiary conditions.

There exists a so-called ``point'' or ``contact'' transformation under which the above Lagrangian is invariant,
\begin{align}
\Psi_\mu&\to\Psi'_\mu=\Psi_\mu+\frac{1}{2}a\gamma_\mu\gamma_\nu\Psi^\nu,\label{pt1}\\
A&\to A'=\frac{A-a}{1+2a},\quad a\neq-\frac{1}{2}.\label{pt2}
\end{align}
The choice for the value of $A$ does not affect physical quantities \cite{Kamefuchi:1961sb,Pilling:2004cu,Krebs:2009bf}. Therefore, one may simplify the above Lagrangian by a field redefinition \cite{Pascalutsa:1994tp}
\begin{eqnarray}
\mathscr{L}_{\mathrm{f}}&=&\bar{\psi}_{A\mu}\Lambda^{\mu\nu}\psi_{A\nu},\label{pt3}\\
\Lambda^{\mu\nu}&=&-(i\slashed{\partial}-m)g^{\mu\nu}+\frac{1}{4}\gamma^\mu\gamma^\lambda(i\slashed{\partial}-m)\gamma_\lambda\gamma^\nu,\nonumber
\end{eqnarray}
where $\psi_A^\mu\equiv O_A^{\mu\nu}\Psi_{\nu}=(g^{\mu\nu}+\frac{1}{2}A\gamma^\mu\gamma^\nu)\Psi_{\nu}$. Now, $\Lambda^{\mu\nu}$ is independent of $A$ and the $A$ dependence is implied in $\psi_{A}^\mu$.

For the meson-decuplet-decuplet (MTT) interactions, the chiral Lagrangian has the form
\begin{eqnarray}
\mathscr{L}_{\mathrm{MTT}}&=&\Tb^{abc}_{\mu}\Lambda_{A,abc}^{def,\mu\nu}T_{def,\nu},\\
\Lambda_{A,abc}^{def,\mu\nu}&=&-\big[(i\slashed{D}-m_T)g^{\mu\nu}+iA(\gamma^\mu D^\nu+\gamma^\nu D^\mu)\nonumber\\
&&+\frac{i}{2}(3A^2+2A+1)\gamma^\mu\slashed{D}\gamma^\nu\nonumber\\
&&+m_T(3A^2+3A+1)\gamma^\mu\gamma^\nu\big]{\delta_{a}}^d{\delta_{b}}^e{\delta_{c}}^f+O_{1,A,abc}^{def,\mu\nu},\label{gcl}
\end{eqnarray}
where $m_T$ is the decuplet mass in the $SU(3)$ limit and $O_{1,A,abc}^{def,\mu\nu}$ contains the meson fields and the external sources. Then the EOM and the subsidiary conditions in ChPT are
\begin{align}
&(i\slashed{D}-m_T)T_{abc}^\mu\doteq 0,\label{eomdi}\\
&D_\mu T_{abc}^\mu\doteq 0,\label{eomdis1}\\
&\gamma_\mu T_{abc}^\mu\doteq 0,\label{eomdis2}
\end{align}
where the symbol ``$\doteq$'' means that both sides are equal if high order terms are ignored. We may write the structure of any term in $O_{1,abc}^{def,\mu\nu}$ as \cite{pin4,Hemmert:1997ye,Jiang:2017yda}
\begin{align}
\Tb^{abc,\mu} O^{\cdots}_{\cdots}\Theta^{\cdots}_{\cdots}T_{def}^\nu+\mathrm{H.c.},\label{form1}
\end{align}
where $\cdots$ denotes suitable flavour and Lorentz indices, $O^{\cdots}_{\cdots}$ is the product of the building blocks with the meson fields and the external sources in Sec. \ref{rev}, and $\Theta^{\cdots}_{\cdots}$ contains a Clifford algebra element $\Gamma\in\{1,\gamma_\mu,\gamf,\fgamma_\mu,\sigma_{\mu\nu}\}$, the Levi-Civita tensors in Lorentz space $\varepsilon^{\mu\nu\lambda\rho}$, and the covariant derivatives acting on $T_{def}^\nu$. Up to the order $\mathcal{O}(p^4)$, the structures of $\Theta^{\cdots}_{\cdots}$ can be found below Eq. (49) in Ref. \cite{Jiang:2017yda}.

With the structure in Eq. \eqref{form1}, the low-energy constants (LECs) in $O_{1,abc}^{def,\mu\nu}$ are dependent on $A$. One can absorb the parameter $A$ into the redefined RS fields according to the point transformation (Eqs. \eqref{pt1} and \eqref{pt3}). Then the Lagrangian \eqref{gcl} can be rewritten as
\begin{align}
\mathscr{L}_{\mathrm{MTT}}&=-\Tb_{A,\mu}^{abc}\big[(i\slashed{D}-m_T)g^{\mu\nu}-\frac{1}{4}\gamma^\mu\gamma^\lambda(i\slashed{D}-m_T)\gamma_\lambda\gamma^\nu\big]\Tb_{A,abc,\nu}+\Tb^{abc}_{A,\mu}O_{1,abc}^{def,\mu\nu}T_{A,def,\nu},\label{gclA}
\end{align}
where $T_{A,abc}^\mu=O_A^{\mu\nu}T_{abc,\nu}$. Now, the LECs in Eq. \eqref{gclA} are independent of $A$, but the invariant monomials have the same structures as those in Eq. \eqref{gcl}, i.e. one may get Eq. \eqref{gclA} from Eq. \eqref{gcl} by changing $T_{abc,\mu}$ to $T_{A,abc,\mu}$ only. The LECs in these two equations are equal if $A=0$. Physically, we can choose any value of $A$ ($A\neq-1/2$) ($A=-1$ is a simple and widely used value). In the final results (Sec. \ref{results}), we only give the structures in Eq. \eqref{gcl}.

The new form $\pi\Delta\Delta$ Lagrangians are very similar to the MTT Lagrangians. The differences lie only in the baryon mass and the flavour indices. By changing $m_T$ to $m_\Delta$ ($\Delta$ mass in the chiral limit) and limiting all the flavour indices to 1 and 2, the new form of $\pi\Delta\Delta$ Lagrangians is obtained.

For the meson-octet-decuplet and $\pi N\Delta$ interactions, the chiral Lagrangians have the following structures, respectively,
\begin{align}
&\epsilon^{abc}{\Bb_d}^e O^{\cdots}_{\cdots}\Theta^{\cdots}_{\cdots}T_{A,n,fgh}^{\mu}+\mathrm{H.c.},\label{form2}\\
&\epsilon^{ab}\psib^c O^{\cdots}_{\cdots}\Theta^{\cdots}_{\cdots}T_{A,n,def}^{\mu}+\mathrm{H.c.},\label{form3}
\end{align}
where $O^{\cdots}_{\cdots}$ and $\Theta^{\cdots}_{\cdots}$ have the same meanings as those in Eq. \eqref{form1}. For the Levi-Civita tensor, we have column indices in $\epsilon^{abc}$ ($a,b,c=1,2,3$) and row indices in $\epsilon_{abc}$ (in the H.c. part). Here, $\epsilon^{ab}\equiv\epsilon^{ab3}$. The RS field depending on $A$ is defined through
\begin{eqnarray}\label{z-para}
T_{A,n,fgh,\mu}&=&\Theta_{A,n,\mu\nu}(z_n)T_{fgh}^{\nu},\\
\Theta_{A,n,\mu\nu}(z_n)&=&g_{\mu\nu}+[z_n+\frac{1}{2}(1+4z_n)A]\gamma_\mu\gamma_\nu\nonumber\\
&\equiv&\Theta_{n,\mu\alpha}(z_n){O_{A}^\alpha}_\nu={O_{A\mu}}^\alpha\Theta_{n,\alpha\nu}(z_n),\notag\\
\Theta_{n,\mu\alpha}(z_n)&\equiv&g_{\mu\alpha}+z_n\gamma_\mu\gamma_\alpha.
\end{eqnarray}
Some $z_n$ parameters are needed because of the point transformation \cite{Nath:1971wp}. They can be obtained from experiments. In Eqs. \eqref{form2} and \eqref{form3}, the point-invariant structures have been implied and the LECs are already independent of $A$.

To construct Lagrangians, for the baryon fields, we choose $T_{abc}^\mu$, $\Tb^{abc,\mu}$, $T_{A,abc}^\mu$, $\Tb_A^{abc,\mu}$, ${B_a}^b$, ${\Bb_a}^b$, and their covariant derivatives as building blocks in the three-flavour case. In the two-flavour case, we adopt $T_{abc}^\mu$, $\Tb^{abc,\mu}$, $T_{A,abc}^\mu$, $\Tb_A^{abc,\mu}$, $\psi_a$, $\psib^a$, and their covariant derivatives.

\section{Preparations for Lagrangian construction}\label{const}

In this section, we make preparations for the construction of chiral Lagrangians with decuplet baryons. The new form of chiral Lagrangians with $\Delta$ is understood. The recipes are very similar to those in constructing Lagrangians for mesons, meson-baryon systems, and the $\pi$-$N$-$\Delta$ systems in Refs. \cite{ourf,Jiang:2016vax,Jiang:2017yda}.

\subsection{Power counting and transformation properties}\label{bbp}

The chiral dimensions \cite{GS1,GS2,Gasser:1987rb,Bijnens:1999sh,pin4,pib31} of the building blocks with the external sources are listed in the second column of Table \ref{blbt} and those of the Clifford algebra and the Levi-Civita tensors are given in the second column of Table \ref{cabt} \cite{pin4,pib31,Scherer:2012xha}. The baryon fields are chiral dimensionless and the information is not shown in these tables. The covariant derivatives acting on the meson fields and the external sources are counted as ${\mathcal O}(p^1)$, but those acting on the baryon fields are counted as ${\mathcal O}(p^0)$.

The chiral Lagrangian should be invariant under the chiral rotation ($R$), parity transformation ($P$), charge conjugation transformation ($C$), and Hermitian transformation (h.c.). The chiral rotations for the building blocks have been discussed in Eqs. \eqref{crm} and \eqref{crmB}. The $P$, $C$, and h.c. transformations are almost the same as those in Ref. \cite{Jiang:2017yda} and we also present such properties in Tables \ref{blbt} and \ref{cabt}. Only different properties will be mentioned.

Compared with Table I of Ref. \cite{Jiang:2017yda}, Table \ref{blbt} here shows the flavour indices explicitly. The meanings of plus and minus signs in Table \ref{cabt} are the same as those in Refs. \cite{pin4,Jiang:2016vax,Jiang:2017yda}. One thing different is the $\epsilon^{ijk}$. This symbol in Ref. \cite{Jiang:2017yda} is in the isovector space and it absorbs a minus sign in $C$ transformations (Eq. (31) of Ref. \cite{Jiang:2017yda}). But now $\epsilon^{abc}$ and $\epsilon^{ab}$ are the Levi-Civita tensors in the three (two)-flavour space. They do not need to absorb an extra minus sign.

\begin{table*}[!h]
\caption{\label{blbt}Chiral dimension (Dim), parity ($P$), charge conjugation ($C$), and Hermiticity (h.c.) of the building blocks with the external sources.}
{\renewcommand\arraystretch{1.3}
\begin{tabular}{ccccc}
	\hline\hline
	                             & Dim &                $P$                 &               $C$                &             h.c.              \\
	\hline
	    $u\ind{_a^{b,\mu}}$      &  1  &         $-u\ind{_a^b_\mu}$         &       $u\ind{_b^{a,\mu}}$        &      $u\ind{_a^{b,\mu}}$      \\
	   $h\ind{_a^{b,\mu\nu}}$    &  2  &      $-h\ind{_a^b_{\mu\nu}}$       &      $h\ind{_b^{a,\mu\nu}}$      &    $h\ind{_a^{b,\mu\nu}}$     \\
	   $\chi\ind{_{\pm,a}^b}$    &  2  &     $\pm\chi\ind{_{\pm,a}^b}$      &      $\chi\ind{_{\pm,b}^a}$      &  $\pm \chi\ind{_{\pm,a}^b}$   \\
	       $\chi_{\pm,s}$        &  2  &         $\pm\chi_{\pm,s}$          &          $\chi_{\pm,s}$          &        $\pm\chi_{\pm,s}$         \\
	$f\ind{_{\pm,a}^{b,\mu\nu}}$ &  2  & $\pm f\ind{_{\pm,a}^{b}_{\mu\nu}}$ & $\mp f\ind{_{\pm,b}^{a,\mu\nu}}$ & $ f\ind{_{\pm,a}^{b,\mu\nu}}$ \\
	  $f\ind{_{+,s}^{\mu\nu}}$   &  2  &       $f\ind{_{+,s,\mu\nu}}$       &    $-f\ind{_{+,s}^{\mu\nu}}$     &   $ f\ind{_{+,s}^{\mu\nu}}$   \\
	\hline\hline
\end{tabular}
}
\end{table*}

\begin{table*}[!h]
\caption{\label{cabt}Chiral dimension (Dim), parity ($P$), charge conjugation ($C$), and Hermiticity (h.c.) of the Clifford algebra elements, the Levi-Civita tensors, and the covariant derivatives. The subscript `$\mathrm{TT}$' (`$\mathrm{BT}$') denotes the meson-decuplet-decuplet (meson-octet-decuplet) interactions in the three flavours ($\pi\Delta\Delta$ ($\pi N\Delta$) interactions in the two-flavour case). $\Psi$ denotes any baryon field, decuplet baryon, $\Delta$, octet baryon, or nucleon. $\epsilon^{abc}$ ($\epsilon^{ab}$) is the Levi-Civita tensor in three (two)-flavour space. The meaning of the plus or minus sign is explained in the text.}
\begin{tabular}{cccccccc}
	\hline\hline
	                               & Dim & $P_\mathrm{TT}$ & $C_\mathrm{TT}$ & h.c.$_\mathrm{TT}$ & $P_\mathrm{BT}$ & $C_\mathrm{BT}$ & h.c.$_\mathrm{BT}$ \\
	\hline
	             $1$               &  0  &   $+$    &   $+$    &     $+$     &   $-$    &   $+$    &     $+$     \\
	           $\gamf$             &  1  &   $-$    &   $+$    &     $-$     &   $+$    &   $+$    &     $-$     \\
	        $\gamma^{\mu}$         &  0  &   $+$    &   $-$    &     $+$     &   $-$    &   $-$    &     $+$     \\
	       $\fgamma^{\mu}$         &  0  &   $-$    &   $+$    &     $+$     &   $+$    &   $+$    &     $+$     \\
	      $\sigma^{\mu\nu}$        &  0  &   $+$    &   $-$    &     $+$     &   $-$    &   $-$    &     $+$     \\
	$\epsilon^{\mu\nu\lambda\rho}$ &  0  &   $-$    &   $+$    &     $+$     &   $-$    &   $+$    &     $+$     \\
	       $\epsilon^{abc}$        &  0  &   $+$    &   $+$    &     $+$     &   $+$    &   $+$    &     $+$     \\
	       $\epsilon^{ab}$         &  0  &   $+$    &   $+$    &     $+$     &   $+$    &   $+$    &     $+$     \\
	         $D^\mu \Psi$          &  0  &   $+$    &   $-$    &     $-$     &   $+$    &   $+$    &     $+$     \\
	\hline\hline
\end{tabular}
\end{table*}

\subsection{Linear relations}\label{lr}
Some linear relations exist in reducing the chiral-invariant terms to a minimal set. The relations coming from partial integration, EOM, covariant derivatives, and Bianchi identity are the same as those in Ref. \cite{Jiang:2017yda}. The relations coming from the Cayley-Hamilton relation are the same as those in Ref. \cite{Bijnens:1999sh}. We will not discuss them any more and we only focus on the different and new relations in the following parts.

\subsubsection{Schouten identity}

The Schouten identity in the Lorentz space is the same as that in Ref. \cite{Jiang:2017yda}, but some differences exist in the flavour space. For the Levi-Civita tensor $\epsilon^{abc}$ ($\epsilon^{ab}$) in the three (two)-flavour space, the Schouten identities for any operator $A$ are
\begin{align}
\begin{split}
0&=\epsilon_{abc}A_d-\epsilon_{dbc}A_a-\epsilon_{adc}A_b-\epsilon_{abd}A_c,\\
0&=\epsilon_{ab}A_c-\epsilon_{cb}A_a-\epsilon_{ac}A_b.
\end{split}\label{si}
\end{align}
There are two types of indices in $A$ (row or column). Eq. \eqref{si} works only for the case that the indices in the Levi-Civita tensor and the indices in $A$ are the same type.

\subsubsection{Fierz transformations}
The basic Fierz transformation for the Pauli matrices is
\begin{align}
\tau^i_{ab}\tau^i_{cd}&=2\delta_{ad}\delta_{cb}-\delta_{ab}\delta_{cd}.
\end{align}
With this equation, for any two $2\times 2$ building blocks ${X_a}^b$ and ${Y_a}^b$ in Table \ref{blbt}, one may obtain \cite{Borodulin:2017pwh}
\begin{align}
X\ind{_a^d}Y\ind{_b^e}=\frac{1}{2}(Y\ind{_a^e}X\ind{_b^d}+X\ind{_a^e}Y\ind{_b^d}+X\ind{_c^f}Y\ind{_f^c}\delta\ind{_a^e}\delta\ind{_b^d}-X\ind{_c^f}Y\ind{_f^c}\delta\ind{_a^d}\delta\ind{_b^e}+X\ind{_a^c}Y\ind{_c^e}\delta\ind{_b^d}-\delta\ind{_a^e}X\ind{_b^c}Y\ind{_c^d}).\label{fierz2}
\end{align}

The basic Fierz transformation for the Gall-Mann matrices is
\begin{align}
\lambda^i_{ac}\lambda^i_{bd}&=2\delta_{ad}\delta_{cb}-\frac{2}{3}\delta_{ac}\delta_{bd}.
\end{align}
With the relation in Ref. \cite{Zong:1994ww} and the properties of the structure constants of $SU(3)$, one finds that the following relation exists for any two $3\times 3$ building blocks ${X_a}^b$ and ${Y_a}^b$ in Table \ref{blbt},
\begin{align}
0={}&X\ind{_a^b}Y\ind{_c^d}-X\ind{_a^d}Y\ind{_c^b}-X\ind{_c^b}Y\ind{_a^d}+X\ind{_c^d}Y\ind{_a^b}+X\ind{_a^e}Y\ind{_e^b}\delta\ind{_c^d}-X\ind{_a^e}Y\ind{_e^d}\delta\ind{_c^b}-X\ind{_c^e}Y\ind{_e^b}\delta\ind{_a^d}+X\ind{_c^e}Y\ind{_e^d}\delta\ind{_a^b}\notag\\
&+\delta\ind{_a^b}Y\ind{_c^e}X\ind{_e^d}-\delta\ind{_a^d}Y\ind{_c^e}X\ind{_e^b}-\delta\ind{_c^b}Y\ind{_a^e}X\ind{_e^d}+\delta\ind{_c^d}Y\ind{_a^e}X\ind{_e^b}-X\ind{_e^f}Y\ind{_f^e}\delta\ind{_a^b}\delta\ind{_c^d}+X\ind{_e^f}Y\ind{_f^e}\delta\ind{_a^d}\delta\ind{_c^b}.\label{fierz3}
\end{align}

\subsubsection{Contact terms}
The method to construct contact terms is the same as that in Ref. \cite{Jiang:2017yda}. In the two (three)-flavour case, the total number of the contact terms is six (five) and we list them in the end of Table \ref{p4pideltab}. The last term in Table \ref{p4pideltab} is at the $\mathcal{O}(p^6)$ order in the $SU(3)$ case.

\subsection{Relations between the original chiral Lagrangians with $\Delta$ and the new ones}\label{rel}
In Ref. \cite{Jiang:2017yda}, we have obtained the chiral Lagrangians with $\Delta$ to one loop. There, the $\Delta$ fields are represented by an isovector-isospinor RS field $\psi_i^\mu$ ($i=1,2,3$). Now, we use a totally symmetrical tensor $T_{abc}^\mu$ ($a,b,c=1,2$) to represent them. The difference lies only in the flavour representations. By some calculations, one gets the following relations between these two formalisms of interaction terms,
\begin{align}
\Tb^{abc}OT_{abc}&=\psib_i O\psi_i,\label{eq37}\\
\Tb^{abe}O\ind{_e^f}T_{abf}&=\psib_iO_j\tau_j\psi_i,\\
\Tb^{abc}X\ind{_b^e}Y\ind{_c^f}T_{aef}&=\psib_i X_jY_j\psi_i-\psib_iX_iY_j\psi_j-\psib_iX_jY_i\psi_j,\\
\Tb^{abc}X\ind{_a^d}Y\ind{_b^e}Z\ind{_c^f}T_{def}&=\frac{1}{6}\psib_iX_lY_j\tau_jZ_l\psi_i-\frac{1}{3}\psib_iX_iY_k\tau_kZ_j\psi_j+P(X,Y,Z),\label{eq40}\\
\epsilon^{ab}\bar{\psi}^{c}O\ind{_a^f}T_{bcf}&=\sqrt{2}\bar{\psi}O_{i}\psi_{i},\label{eq41}\\
\epsilon^{ab}\bar{\psi}^{c}X\ind{_a^e}Y\ind{_c^f}T_{ebf}&=\sqrt{2}\bar{\psi}X_{i}Y_{j}\tau_{j}\psi_{i},\\
\epsilon^{ab}\bar{\psi}^{c}X\ind{_a^e}Y\ind{_b^f}T_{efc}&=\sqrt{2}i\epsilon_{ijk}\bar{\psi}X_{i}Y_{j}\psi_{k},\\
\epsilon^{ab}\bar{\psi}^{c}X\ind{_a^d}Y\ind{_b^e}Z\ind{_c^f}T_{def}&=\sqrt{2}i\epsilon_{ijk}\bar{\psi}X_{i}Y_{j}Z_{l}\tau_{l}\psi_{k},\label{eq44}
\end{align}
where $P(X,Y,Z)$ means all permutations for the symbols $X$, $Y$, and $Z$. $O$, $O_i$, $X_i$, $Y_i$ and $Z_i$ are building blocks in Ref. \cite{Jiang:2017yda} or their products. The definitions of the symbols in the right-hand side can be found in Ref. \cite{Jiang:2017yda}.

Alternatively, we may transform the original formalism to the new one. To do that, we define transition isospin $I^j$ through $\psi_j=I_j\phi$ with $\phi=(\Delta^{++},\Delta^+,\Delta^0,\Delta^-)^T$. Similarly, we define $T_{abc}=W_{abc}^i\phi^i$. The matrix forms of $I^j$ and the values of $W_{abc}^i$ are easy to obtain from the definitions. We have two relations in connecting the original $\pi\Delta\Delta$ Lagrangians with the new ones,
\begin{eqnarray}
{(I_i^\dag I_j)_x}^y
&=&\frac12[W^{abc}_x{(\tau_i\tau_j)_a}^dW_{bcd}^y- W^{abc}_x{(\tau_i)_a}^d{(\tau_j)_b}^eW_{cde}^y],\\
{(I_i^\dag \tau_lI_j)_x}^y
&=&\frac12[W^{abc}_x{(\tau_i\tau_j)_a}^d{(\tau_l)_b}^eW_{cde}^y- W^{abc}_x{(\tau_i)_a}^d{(\tau_j)_b}^e{(\tau_l)_c}^fW_{def}^y].
\end{eqnarray}
For the special case $j=i$, one has
\begin{eqnarray}
{(I_i^\dag I_i)_x}^y&=&W^{abc}_xW_{abc}^y,\\
{(I_i^\dag \tau_lI_i)_x}^y&=&W^{abc}_x{(\tau_l)_a}^dW_{bcd}^y.
\end{eqnarray}
To connect the original $\pi N\Delta$ Lagrangians with the new ones, we may use
\begin{eqnarray}
{(I_i)_x}^y&=&\frac{1}{\sqrt2}\epsilon^{3ab}{(\tau_i)_a}^cW_{xbc}^y,\\
{(\tau_iI_j)_x}^y&=&\frac{1}{\sqrt2}\epsilon^{3ab}{(\tau_i)_x}^c{(\tau_j)_a}^dW^y_{bcd}.
\end{eqnarray}
Note ${(\tau_iI_j)_x}^y\neq\frac{1}{\sqrt2}\epsilon^{3ab}{(\tau_i)_a}^c{(\tau_j)_x}^dW^y_{bcd}$. Substituting these six equations into the right-hand sides of Eqs. \eqref{eq37}-\eqref{eq44}, one may prove the equivalence of the two sets of relations by using the formula $\epsilon^{3ab}{(\tau_i\tau_j)_a}^cW_{xbc}^y=\epsilon^{3ab}{(\tau_i)_a}^c{(\tau_j)_b}^dW_{xcd}^y$.

\section{Results and discussions}\label{results}

Following the same steps from Sec. IV C to Sec. IV E in Ref. \cite{Jiang:2017yda}, we obtain the chiral Lagrangians with decuplet baryons up to the order ${\mathcal O}(p^4)$ and list them below.

\subsection{$\mathcal{O}(p^1)$ order}

In the three-flavour case, the lowest order meson-decuplet-decuplet chiral Lagrangian is
\begin{align}\label{p1pid}
\mathscr{L}^{(1)}_\mathrm{MTT}&=\cdots+C^{(1)}_1\Tb^{ a b c\mu}{u_{ a}}^{ d\nu}\gamf\gamma_{\nu}T_{ b c d\mu},
\end{align}
where $C^{(1)}_1$ is the low-energy constant at this order and the ellipsis represents the terms coming from the first part in Eq. \eqref{gcl}. The lowest order meson-octet-decuplet chiral Lagrangian reads
\begin{align}
\mathscr{L}^{(1)}_\mathrm{MBT}&=D^{(1)}_1\epsilon^{ a b c}{\Bb_{ a}}^{ d}{u_{ b}}^{ e\mu}T_{ c d e\mu}+\mathrm{H.c.}.
\end{align}

In the two-flavour case, the lowest order $\pi\Delta\Delta$ chiral Lagrangian has the same form as Eq. \eqref{p1pid},
\begin{align}
\mathscr{L}^{(1)}_{\pi\Delta\Delta}&=\cdots+e^{(1)}_1\Tb^{ a b c\mu}{u_{ a}}^{ d\nu}\gamf\gamma_{\nu}T_{ b c d\mu}.
\end{align}
The difference lies only in the allowed numbers for the indices $a$, $b$, $c$, and $d$. Similarly, the lowest order $\pi N\Delta$ chiral Lagrangian can be written as
\begin{align}
\mathscr{L}^{(1)}_{\pi N\Delta}&=f^{(1)}_1(\epsilon^{ a b}\psib^{ c}{u_{ a}}^{ d\mu}T_{A,n, b c d\mu}+\mathrm{H.c.}).
\end{align}
We have confirmed the previous results in Ref. \cite{Jiang:2017yda} with the newly constructed Lagrangians. With the relations in the last section, we get the relations between these two kinds of LECs,
\begin{align}
e^{(1)}_1=c_1^{(1)}=\frac{1}{2}g_1,\;f^{(1)}_1=\frac{1}{\sqrt{2}}g_{\pi N\Delta}
\end{align}

\subsection{$\mathcal{O}(p^2)$ order}
The $\mathcal{O}(p^2)$ order meson-decuplet-decuplet chiral Lagrangian has the form
\begin{align}
\mathscr{L}^{(2)}_\mathrm{MTT}=\sum_{n=1}^{13}C^{(2)}_n O^{(3,2)}_n,
\end{align}
where the operators $O^{(N_f=3,2)}_n$ are listed in Table \ref{p2pideltab}. The meson-octet-decuplet chiral Lagrangian at this order is
\begin{align}
\mathscr{L}^{(2)}_\mathrm{MBT}=&D^{(2)}_1(\epsilon^{ a b c}{\Bb_{ a}}^{ d}{u_{ b}}^{ e\mu}{u_{ c}}^{ f\nu}\gamf\gamma_{\mu}T_{A,n, d e f\nu}+\mathrm{H.c.})
+D^{(2)}_2(\epsilon^{ a b c}{\Bb_{ a}}^{ d}{u_{ b}}^{ e\mu}{u_{ d}}^{ f\nu}\gamf\gamma_{\mu}T_{A,n, c e f\nu}+\mathrm{H.c.})\notag\\
&+D^{(2)}_3(\epsilon^{ a b c}{\Bb_{ a}}^{ d}{u_{ b}}^{ e\mu}{u_{ e}}^{ f\nu}\gamf\gamma_{\mu}T_{A,n, c d f\nu}+\mathrm{H.c.})
+D^{(2)}_4(\epsilon^{ a b c}{\Bb_{ a}}^{ d}{u_{ b}}^{ e\mu}{u_{ e}}^{ f\nu}\gamf\gamma_{\nu}T_{A,n, c d f\mu}+\mathrm{H.c.})\notag\\
&+D^{(2)}_5(i\epsilon^{ a b c}{\Bb_{ a}}^{ d}{f_{+ b}}^{ e\mu\nu}\gamf\gamma_{\mu}T_{A,n, c d e\nu}
+\mathrm{H.c.}).
\end{align}
This result is consistent with that in Ref. \cite{Li:2017vmq}.

\begin{table*}[!h]
\caption{\label{p2pideltab}The order $\mathcal{O}(p^2)$ meson-decuplet-decuplet ($\pi\Delta\Delta$) chiral Lagrangians, and the relations between $\pi\Delta\Delta$ LECs here and those in Ref. \cite{Jiang:2017yda}.}
\begin{tabular}{lccl}
	\hline\hline
	$O^{(n_f,2)}_n$                                                                  & $SU(2)$ & $SU(3)$ & $e^{(2)}_n$                \\
	\hline
	$\Tb^{ a b c\mu}{{u_{ a}}^{ d}}_{\mu}{u_{ b}}^{ e\nu}T_{ c d e\nu}$              &    1    &    1    & $-c^{(2)}_1/2-c^{(2)}_2/2$ \\
	$\Tb^{ a b c\mu}{u_{ a}}^{ d\nu}{{u_{ b}}^{ e}}_{\nu}T_{ c d e\mu}$              &    2    &    2    & $-c^{(2)}_3/2$             \\
	$\Tb^{ a b c\mu}{{u_{ a}}^{ d}}_{\mu}{u_{ d}}^{ e\nu}T_{ b c e\nu}$              &    3    &    3    & $c^{(2)}_1/2+c^{(2)}_4/2$  \\
	$\Tb^{ a b c\mu}{u_{ a}}^{ d\nu}{{u_{ d}}^{ e}}_{\mu}T_{ b c e\nu}$              &    4    &    4    & $c^{(2)}_2/2+c^{(2)}_4/2$  \\
	$\Tb^{ a b c\mu}{u_{ a}}^{ d\nu}{{u_{ d}}^{ e}}_{\nu}T_{ b c e\mu}$              &    5    &    5    & $c^{(2)}_3/2+c^{(2)}_5$    \\
	$\Tb^{ a b c\mu}{u^{ d e}}_{\mu}{u_{ e d}}^{\nu}T_{ a b c\nu}$                   &         &    6    &                            \\
	$\Tb^{ a b c\mu}u^{ d e\nu}u_{ e d\nu}T_{ a b c\mu}$                             &         &    7    &                            \\
	$\Tb^{ a b c\mu}{u_{ a}}^{ d\nu}{u_{ b}}^{ e\lambda}D_{\nu\lambda}T_{ c d e\mu}$ &    6    &    8    & $-c^{(2)}_6/2$             \\
	$\Tb^{ a b c\mu}{u_{ a}}^{ d\nu}{u_{ d}}^{ e\lambda}D_{\nu\lambda}T_{ b c e\mu}$ &    7    &    9    & $c^{(2)}_6/2+c^{(2)}_7$    \\
	$\Tb^{ a b c\mu}u^{ d e\nu}{u_{ e d}}^{\lambda}D_{\nu\lambda}T_{ a b c\mu}$      &         &   10    &                            \\
	$i\Tb^{ a b c\mu}{f_{s,+\mu}}^{\nu}T_{ a b c\nu}$                                &    8   &         & $c^{(2)}_8$                \\
	$i\Tb^{ a b c\mu}{{{f_{+ a}}^{ d}}_{\mu}}^{\nu}T_{ b c d\nu}$                    &    9    &   11    & $c^{(2)}_9$                \\
	$\Tb^{ a b c\mu}\chi_{+,s} T_{ a b c\mu}$                                        &   10    &   12    & $c^{(2)}_{10}$             \\
	$\Tb^{ a b c\mu}{\chi_{+ a}}^{ d}T_{ b c d\mu}$                                  &   11    &   13    & $c^{(2)}_{11}$             \\
	\hline\hline
\end{tabular}
\end{table*}

The new form of the $\pi\Delta\Delta$ chiral Lagrangian at the $\mathcal{O}(p^2)$ order is
\begin{align}
\mathscr{L}^{(2)}_\mathrm{\pi\Delta\Delta}=\sum_{n=1}^{11}e^{(2)}_n O^{(2,2)}_n,
\end{align}
where the operators $O^{(N_f=2,2)}_n$ can also be found in Table \ref{p2pideltab}. The new form $\pi N\Delta$ chiral Lagrangian reads
\begin{align}
\mathscr{L}^{(2)}_{\pi N\Delta}=&f^{(2)}_1(\epsilon^{ a b}\psib^{ c}{u_{ a}}^{ d\mu}{u_{ b}}^{ e\nu}\gamf\gamma_{\mu}T_{A,n, c d e\nu}+\mathrm{H.c.})
+f^{(2)}_2(\epsilon^{ a b}\psib^{ c}{u_{ a}}^{ d\mu}{u_{ c}}^{ e\nu}\gamf\gamma_{\mu}T_{A,n, b d e\nu}+\mathrm{H.c.})\notag\\
&+f^{(2)}_3(i\epsilon^{ a b}\psib^{ c}{f_{+ a}}^{ d\mu\nu}\gamf\gamma_{\mu}T_{A,n, b c d\nu}
+\mathrm{H.c.}).
\end{align}
This result is consistent with the Lagrangian in Ref. \cite{Jiang:2017yda}. We present the relations between these two kinds of $\pi\Delta\Delta$ LECs in the last column of Table \ref{p2pideltab}. The obtained relations for the $\pi N\Delta$ LECs are
\begin{align}
f^{(2)}_1&=-\frac{1}{\sqrt{2}}d^{(2)}_1,\\
f^{(2)}_2&=\frac{1}{\sqrt{2}}d^{(2)}_1+\frac{1}{\sqrt{2}}d^{(2)}_2,\\
f^{(2)}_3&=\frac{1}{\sqrt{2}}d^{(2)}_3.
\end{align}

\subsection{$\mathcal{O}(p^3)$ and $\mathcal{O}(p^4)$ orders}

We define the $\mathcal{O}(p^3)$ and $\mathcal{O}(p^4)$ chiral Lagrangians as
\begin{eqnarray}
\mathscr{L}^{(m)}_\mathrm{MTT}&=&\sum_{n}C^{(m)}_n O^{(3,m)}_n,\label{pidecn}\\
\mathscr{L}^{(m)}_\mathrm{MBT}&=&\sum_{n}D^{(m)}_n (P^{(3,m)}_n+\mathrm{H.c.}),\label{pindecn}\\
\mathscr{L}^{(m)}_{\pi\Delta\Delta}&=&\sum_{n}e^{(m)}_n O^{(2,m)}_n,\label{pideln}\\
\mathscr{L}^{(m)}_{\pi N\Delta}&=&\sum_{n}f^{(m)}_n (P^{(2,m)}_n+\mathrm{H.c.}),\label{pindeln}
\end{eqnarray}
where $m=3$ or 4 denotes the chiral dimension, $C^{(m)}_n$, $D^{(m)}_n$, $e^{(m)}_n$, and $f^{(m)}_n$ are the LECs, and $O^{(N_f,m)}_n$ and $P^{(N_f,m)}_n$ are the independent chiral-invariant terms in the $N_f$-flavour case. The results are listed in Appendix \ref{aeom}. At the $\mathcal{O}(p^3)$ order, the meson-decuplet-decuplet ($\pi\Delta\Delta$) Lagrangians are presented in Table \ref{p3pideltab}. There are 55 (38) independent terms in the $SU(3)$ ($SU(2)$) case. The meson-octet-decuplet ($\pi N\Delta$) Lagrangians are given in Table \ref{p3pindeltab}. There are 67 (33) independent terms in the $SU(3)$ ($SU(2)$) case. At the $\mathcal{O}(p^4)$ order, the meson-decuplet-decuplet ($\pi\Delta\Delta$) Lagrangians are presented in Table \ref{p4pideltab}. There are 548 (318) independent terms in the $SU(3)$ ($SU(2)$) case. The meson-octet-decuplet ($\pi N\Delta$) Lagrangians are listed in Table \ref{p4pindeltab}. There are 611 (218) independent terms in the $SU(3)$ ($SU(2)$) case. Note that the $z_n$ parameters should be different for the meson-octet-decuplet and $\pi N\Delta$ Lagrangians at the different orders, but we do not distinguish them explicitly in the results.

To merge the meson-octet-decuplet and the $\pi N\Delta$ results, similar to those for the meson-decuplet-decuplet and $\pi\Delta\Delta$, we write them in a unified form. We have changed the $SU(2)$ results with $\epsilon^{ab}\psib^{c}\to\epsilon^{dab}{\Bb_d}^{c}$ by setting $d=3$ but $a,b,c=1,2$ as before. Now, one can get the $SU(2)$ results from corresponding terms in Table \ref{p3pindeltab} and Table \ref{p4pindeltab} with
\begin{align}
\epsilon^{abc}{\Bb_a}^d\cdots\xrightarrow{N_f=2}\epsilon^{bc}\psib^d\cdots.\label{fors}
\end{align}

Because the number of LECs in $\mathcal{O}(p^3)$ and $\mathcal{O}(p^4)$ Lagrangians is large and only several LECs will be involved in a study, we here do not give the LEC relations between the new and original results at high orders. Each form of Lagrangians can be chosen to study low-energy processes. One may use relations in Sec. \ref{rel} to determine LECs from another form terms, if necessary.

From the results, one can see that not only the total number of terms but also the number in each type of external sources in the chiral Lagrangians with $\Delta$ are the same as those in Ref. \cite{Jiang:2017yda}. The equality in number is a strict condition for consistency of Lagrangians in different forms. The violation of this condition means that the number of terms in either or both forms is not minimal. This check confirms our previous results.

\section{Summary}\label{summ}

In this paper, we construct the relativistic chiral Lagrangians with decuplet baryons and give a new form of the chiral Lagrangians with $\Delta(1232)$ to one loop. These chiral Lagrangians are for the meson-decuplet-decuplet, meson-octet-decuplet, $\pi\Delta\Delta$, and $\pi N\Delta$ interactions. The correspondence between the $\pi\Delta\Delta$ and $\pi N\Delta$ chiral Lagrangians in Ref. \cite{Jiang:2017yda} and those in the present form can be obtained with the relations we get in Sec. \ref{rel}. 

\section*{Acknowledgements}

We thank Prof. Li-Sheng Geng for useful discussions. YRL also thanks the hospitality from Prof. M. Oka and other people at Tokyo Institute of Technology where the draft was finished. This work was supported by the National Science Foundation of China (NSFC) under Grants No. 11565004, No. 11775132, No. U1731239, and No. 11673006, the special funding for Guangxi distinguished professors (Bagui Yingcai and Bagui Xuezhe) and High Level Innovation Team and Outstanding Scholar Program in Guangxi Colleges.

\appendix
\section{Independent terms in $\mathcal{O}(p^3)$ and $\mathcal{O}(p^4)$ chiral Lagrangians with decuplet baryons}\label{aeom}

\begin{longtable}{lcclcc}
\caption{Terms in the $\mathcal{O}(p^3)$ meson-decuplet-decuplet and $\pi \Delta\Delta$ chiral Lagrangians, where $O^{(N_f,3)}_n$ is defined in Eqs. \eqref{pidecn} and \eqref{pideln}.}\label{p3pideltab}\\
\hline\hline $O^{(N_f,3)}_n$ & $SU(2)$ &  $SU(3)$ & $O^{(N_f,3)}_n$ & $SU(2)$ &  $SU(3)$\\
\hline\endfirsthead

\hline\hline $O^{(N_f,3)}_n$ & $SU(2)$ &  $SU(3)$ & $O^{(N_f,3)}_n$ & $SU(2)$ &  $SU(3)$\\
\hline\endhead

\hline\hline 
\endfoot

\hline\endlastfoot

$\Tb^{ a b c\mu}{{u_{ a}}^{ d}}_{\mu}{u_{ b}}^{ e\nu}{u_{ c}}^{ f\lambda}\gamf\gamma_{\nu}T_{ d e f\lambda}$ & 1 & 1 & $\Tb^{ a b c\mu}{u_{ a}}^{ d\nu}{{{f_{- d}}^{ e}}_{\mu}}^{\lambda}D_{\nu}T_{ b c e\lambda}+\mathrm{H.c.}$ & 15 & 30  \\
$\Tb^{ a b c\mu}{{u_{ a}}^{ d}}_{\mu}{u_{ b}}^{ e\nu}{u_{ d}}^{ f\lambda}\gamf\gamma_{\nu}T_{ c e f\lambda}$ & 2 & 2 & $\Tb^{ a b c\mu}{u_{ a}}^{ d\nu}{{{f_{- d}}^{ e}}_{\mu}}^{\lambda}D_{\lambda}T_{ b c e\nu}+\mathrm{H.c.}$ & 16 & 31  \\
$\Tb^{ a b c\mu}{{u_{ a}}^{ d}}_{\mu}{u_{ b}}^{ e\nu}{u_{ d}}^{ f\lambda}\gamf\gamma_{\lambda}T_{ c e f\nu}+\mathrm{H.c.}$ & 3 & 3 & $\Tb^{ a b c\mu}{u_{ a}}^{ d\nu}{{{f_{- d}}^{ e}}_{\nu}}^{\lambda}D_{\lambda}T_{ b c e\mu}+\mathrm{H.c.}$ & 17 & 32  \\
$\Tb^{ a b c\mu}{{u_{ a}}^{ d}}_{\mu}{u_{ b}}^{ e\nu}{u_{ e}}^{ f\lambda}\gamf\gamma_{\lambda}T_{ c d f\nu}+\mathrm{H.c.}$ & 4 & 4 & $\Tb^{ a b c\mu}{u^{ d e}}_{\mu}{f_{- e d}}^{\nu\lambda}D_{\nu}T_{ a b c\lambda}+\mathrm{H.c.}$ &  & 33  \\
$\Tb^{ a b c\mu}{u_{ a}}^{ d\nu}{{u_{ b}}^{ e}}_{\nu}{u_{ c}}^{ f\lambda}\gamf\gamma_{\lambda}T_{ d e f\mu}$ & 5 & 5 & $\Tb^{ a b c\mu}u^{ d e\nu}{f_{- e d\mu}}^{\lambda}D_{\nu}T_{ a b c\lambda}$ &  & 34  \\
$\Tb^{ a b c\mu}{{u_{ a}}^{ d}}_{\mu}{u_{ d}}^{ e\nu}{u_{ e}}^{ f\lambda}\gamf\gamma_{\nu}T_{ b c f\lambda}$ &  & 6 & $\Tb^{ a b c\mu}{{u_{ a}}^{ d}}_{\mu}{h_{ d}}^{ e\nu\lambda}D_{\nu}T_{ b c e\lambda}+\mathrm{H.c.}$ & 18 & 35  \\
$\Tb^{ a b c\mu}{u_{ a}}^{ d\nu}{{u_{ b}}^{ e}}_{\nu}{u_{ d}}^{ f\lambda}\gamf\gamma_{\lambda}T_{ c e f\mu}+\mathrm{H.c.}$ & 6 & 7 & $\Tb^{ a b c\mu}{u_{ a}}^{ d\nu}{{{h_{ d}}^{ e}}_{\nu}}^{\lambda}D_{\lambda}T_{ b c e\mu}+\mathrm{H.c.}$ & 19 & 36  \\
$\Tb^{ a b c\mu}{u_{ a}}^{ d\nu}{u_{ b}}^{ e\lambda}{{u_{ d}}^{ f}}_{\mu}\gamf\gamma_{\lambda}T_{ c e f\nu}$ & 7 & 8 & $\Tb^{ a b c\mu}{u_{ a}}^{ d\nu}{h_{ d}}^{ e\lambda\rho}D_{\nu\lambda\rho}T_{ b c e\mu}+\mathrm{H.c.}$ & 20 & 37  \\
\hline\hline
\end{longtable}

\begin{longtable}{lcclcc}
\caption{\label{p4pideltab}Terms in the $\mathcal{O}(p^4)$ meson-decuplet-decuplet and $\pi \Delta\Delta$ chiral Lagrangians, where $O^{(N_f,4)}_n$ is defined in Eqs. \eqref{pidecn} and \eqref{pideln}.}\\
\hline\hline $O^{(N_f,4)}_n$ & $SU(2)$ &  $SU(3)$ & $O^{(N_f,4)}_n$ & $SU(2)$ &  $SU(3)$\\
\hline\endfirsthead

\hline\hline $O^{(N_f,4)}_n$ & $SU(2)$ &  $SU(3)$ & $O^{(N_f,4)}_n$ & $SU(2)$ &  $SU(3)$\\
\hline\endhead

\hline\hline 
\endfoot

\hline\endlastfoot

$\Tb^{ a b c\mu}{{u_{ a}}^{ d}}_{\mu}{u_{ b}}^{ e\nu}{{u_{ c}}^{ f}}_{\nu}{u_{ d}}^{ g\lambda}T_{ e f g\lambda}$ & 1 & 1 & $\Tb^{ a b c\mu}h^{ d e\nu\lambda}{h_{ e d}}^{\rho\sigma}D_{\nu\lambda\rho\sigma}T_{ a b c\mu}$ &  & 299  \\
$\Tb^{ a b c\mu}{{u_{ a}}^{ d}}_{\mu}{u_{ b}}^{ e\nu}{{u_{ c}}^{ f}}_{\nu}{u_{ e}}^{ g\lambda}T_{ d f g\lambda}+\mathrm{H.c.}$ & 2 & 2 & $i\Tb^{ a b c\mu}{{{h_{ a}}^{ d}}_{\mu}}^{\nu}{h_{ d}}^{ e\lambda\rho}\sigma_{\nu\lambda}T_{ b c e\rho}$ & 122 & 300  \\
$\Tb^{ a b c\mu}{{u_{ a}}^{ d}}_{\mu}{u_{ b}}^{ e\nu}{u_{ c}}^{ f\lambda}{{u_{ e}}^{ g}}_{\nu}T_{ d f g\lambda}$ & 3 & 3 & $\Tb^{ a b c\mu}{{u_{ a}}^{ d}}_{\mu}\nabla^{\nu}{{{f_{- b}}^{ e}}_{\nu}}^{\lambda}T_{ c d e\lambda}+\mathrm{H.c.}$ & 123 & 301  \\
$\Tb^{ a b c\mu}{{u_{ a}}^{ d}}_{\mu}{u_{ b}}^{ e\nu}{u_{ c}}^{ f\lambda}{{u_{ e}}^{ g}}_{\lambda}T_{ d f g\nu}+\mathrm{H.c.}$ & 4 & 4 & $\Tb^{ a b c\mu}{u_{ a}}^{ d\nu}\nabla^{\lambda}{{f_{- b}}^{ e}}_{\nu\lambda}T_{ c d e\mu}$ & 124 & 302  \\
$\Tb^{ a b c\mu}{{u_{ a}}^{ d}}_{\mu}{u_{ b}}^{ e\nu}{{u_{ d}}^{ f}}_{\nu}{u_{ e}}^{ g\lambda}T_{ c f g\lambda}$ & 5 & 5 & $\Tb^{ a b c\mu}{{u_{ a}}^{ d}}_{\mu}\nabla^{\nu}{{{f_{- d}}^{ e}}_{\nu}}^{\lambda}T_{ b c e\lambda}+\mathrm{H.c.}$ & 125 & 303  \\
$\Tb^{ a b c\mu}{{u_{ a}}^{ d}}_{\mu}{u_{ b}}^{ e\nu}{u_{ d}}^{ f\lambda}{{u_{ e}}^{ g}}_{\nu}T_{ c f g\lambda}$ &  & 6 & $\Tb^{ a b c\mu}{u_{ a}}^{ d\nu}\nabla^{\lambda}{{f_{- d}}^{ e}}_{\mu\lambda}T_{ b c e\nu}+\mathrm{H.c.}$ & 126 & 304  \\
$\Tb^{ a b c\mu}{{u_{ a}}^{ d}}_{\mu}{u_{ b}}^{ e\nu}{{u_{ d}}^{ f}}_{\nu}{u_{ f}}^{ g\lambda}T_{ c e g\lambda}$ &  & 7 & $\Tb^{ a b c\mu}{u_{ a}}^{ d\nu}\nabla^{\lambda}{{f_{- d}}^{ e}}_{\nu\lambda}T_{ b c e\mu}+\mathrm{H.c.}$ & 127 & 305  \\
$\Tb^{ a b c\mu}{{u_{ a}}^{ d}}_{\mu}{u_{ b}}^{ e\nu}{u_{ d}}^{ f\lambda}{{u_{ e}}^{ g}}_{\lambda}T_{ c f g\nu}+\mathrm{H.c.}$ & 6 & 8 & $\Tb^{ a b c\mu}{u^{ d e}}_{\mu}\nabla^{\nu}{f_{- e d\nu}}^{\lambda}T_{ a b c\lambda}+\mathrm{H.c.}$ &  & 306  \\
\hline\hline
\end{longtable}

\begin{longtable}{lcclcc}
\caption{\label{p3pindeltab}Terms in the $\mathcal{O}(p^3)$ meson-octet-decuplet and $\pi N\Delta$ chiral Lagrangians, where $P^{(N_f,3)}_n$ is defined in Eqs. \eqref{pindecn} and \eqref{pindeln}. For the $SU(2)$ case, the form needs to be changed, see the sentences around Eq. \eqref{fors}.}\\
\hline\hline $P^{(N_f,3)}_n$ & $SU(2)$ & $SU(3)$ & $P^{(N_f,3)}_n$ & $SU(2)$ & $SU(3)$\\
\hline\endfirsthead

\hline\hline $P^{(N_f,3)}_n$ & $SU(2)$ & $SU(3)$ & $P^{(N_f,3)}_n$ & $SU(2)$ & $SU(3)$\\
\hline\endhead

\hline\hline 
\endfoot

\hline\endlastfoot

$\epsilon^{ a b c}{\Bb_{ a}}^{ d}{u_{ b}}^{ e\mu}{{u_{ c}}^{ f}}_{\mu}{u_{ e}}^{ g\nu}T_{A,n, d f g\nu}$ & 1 & 1 & $\epsilon^{ a b c}{\Bb_{ a}}^{ d}{u_{ b}}^{ e\mu}{f_{- e}}^{ f\nu\lambda}\gamf\gamma_{\nu}D_{\lambda}T_{A,n, c d f\mu}$ &  & 36  \\
$\epsilon^{ a b c}{\Bb_{ a}}^{ d}{u_{ b}}^{ e\mu}{u_{ c}}^{ f\nu}{{u_{ d}}^{ g}}_{\mu}T_{A,n, e f g\nu}$ & 2 & 2 & $\epsilon^{ a b c}{\Bb_{ a}}^{ d}u^{ e f\mu}{f_{- b e}}^{\nu\lambda}\gamf\gamma_{\mu}D_{\nu}T_{A,n, c d f\lambda}$ &  & 37  \\
$\epsilon^{ a b c}{\Bb_{ a}}^{ d}{u_{ b}}^{ e\mu}{u_{ c}}^{ f\nu}{{u_{ e}}^{ g}}_{\mu}T_{A,n, d f g\nu}$ &  & 3 & $\epsilon^{ a b c}{\Bb_{ a}}^{ d}u^{ e f\mu}{f_{- b e}}^{\nu\lambda}\gamf\gamma_{\nu}D_{\mu}T_{A,n, c d f\lambda}$ &  & 38  \\
$\epsilon^{ a b c}{\Bb_{ a}}^{ d}{u_{ b}}^{ e\mu}{u_{ c}}^{ f\nu}{{u_{ e}}^{ g}}_{\nu}T_{A,n, d f g\mu}$ & 3 & 4 & $\epsilon^{ a b c}{\Bb_{ a}}^{ d}u^{ e f\mu}{f_{- b e}}^{\nu\lambda}\gamf\gamma_{\nu}D_{\lambda}T_{A,n, c d f\mu}$ &  & 39  \\
$\epsilon^{ a b c}{\Bb_{ a}}^{ d}{u_{ b}}^{ e\mu}{{u_{ d}}^{ f}}_{\mu}{u_{ e}}^{ g\nu}T_{A,n, c f g\nu}$ &  & 5 & $\epsilon^{ a b c}{\Bb_{ a}}^{ d}{u_{ b}}^{ e\mu}{h_{ c}}^{ f\nu\lambda}\gamf\gamma_{\mu}D_{\nu}T_{A,n, d e f\lambda}$ & 15 & 40  \\
$\epsilon^{ a b c}{\Bb_{ a}}^{ d}{u_{ b}}^{ e\mu}{u_{ d}}^{ f\nu}{{u_{ e}}^{ g}}_{\mu}T_{A,n, c f g\nu}$ &  & 6 & $\epsilon^{ a b c}{\Bb_{ a}}^{ d}{u_{ b}}^{ e\mu}{h_{ d}}^{ f\nu\lambda}\gamf\gamma_{\mu}D_{\nu}T_{A,n, c e f\lambda}$ & 16 & 41  \\
$\epsilon^{ a b c}{\Bb_{ a}}^{ d}{u_{ b}}^{ e\mu}{u_{ d}}^{ f\nu}{{u_{ e}}^{ g}}_{\nu}T_{A,n, c f g\mu}$ &  & 7 & $\epsilon^{ a b c}{\Bb_{ a}}^{ d}{u_{ b}}^{ e\mu}{h_{ e}}^{ f\nu\lambda}\gamf\gamma_{\mu}D_{\nu}T_{A,n, c d f\lambda}$ &  & 42  \\
$\epsilon^{ a b c}{\Bb_{ a}}^{ d}{u_{ b}}^{ e\mu}{{u_{ e}}^{ f}}_{\mu}{u_{ f}}^{ g\nu}T_{A,n, c d g\nu}$ &  & 8 & $\epsilon^{ a b c}{\Bb_{ a}}^{ d}{u_{ b}}^{ e\mu}{h_{ e}}^{ f\nu\lambda}\gamf\gamma_{\nu}D_{\mu}T_{A,n, c d f\lambda}$ &  & 43  \\
\hline\hline
\end{longtable}

\begin{longtable}{lcclcc}
\caption{\label{p4pindeltab}Terms in the $\mathcal{O}(p^4)$ meson-octet-decuplet and $\pi N\Delta$ chiral Lagrangians, where $P^{(N_f,4)}_n$ is defined in Eqs. \eqref{pindecn} and \eqref{pindeln}. For the $SU(2)$ case, the form needs to be changed, see the sentences around Eq. \eqref{fors}.}\\
\hline\hline $P^{(N_f,4)}_n$ & $SU(2)$ & $SU(3)$ & $P^{(N_f,4)}_n$ & $SU(2)$ & $SU(3)$\\
\hline\endfirsthead

\hline\hline $P^{(N_f,4)}_n$ & $SU(2)$ & $SU(3)$ & $P^{(N_f,4)}_n$ & $SU(2)$ & $SU(3)$\\
\hline\endhead

\hline\hline 
\endfoot

\hline\endlastfoot

$\epsilon^{ a b c}{\Bb_{ a}}^{ d}{u_{ b}}^{ e\mu}{{u_{ c}}^{ f}}_{\mu}{u_{ d}}^{ g\nu}{u_{ e}}^{ h\lambda}\gamf\gamma_{\nu}T_{A,n, f g h\lambda}$ & 1 & 1 & $\epsilon^{ a b c}{\Bb_{ a}}^{ d}{u_{ b}}^{ e\mu}\nabla^{\nu}{{{f_{- c}}^{ f}}_{\mu}}^{\lambda}\gamf\gamma_{\nu}T_{A,n, d e f\lambda}$ & 78 & 322  \\
$\epsilon^{ a b c}{\Bb_{ a}}^{ d}{u_{ b}}^{ e\mu}{{u_{ c}}^{ f}}_{\mu}{u_{ d}}^{ g\nu}{u_{ e}}^{ h\lambda}\gamf\gamma_{\lambda}T_{A,n, f g h\nu}$ & 2 & 2 & $\epsilon^{ a b c}{\Bb_{ a}}^{ d}{u_{ b}}^{ e\mu}\nabla^{\nu}{{{f_{- c}}^{ f}}_{\nu}}^{\lambda}\gamf\gamma_{\mu}T_{A,n, d e f\lambda}$ & 79 & 323  \\
$\epsilon^{ a b c}{\Bb_{ a}}^{ d}{u_{ b}}^{ e\mu}{{u_{ c}}^{ f}}_{\mu}{u_{ e}}^{ g\nu}{u_{ f}}^{ h\lambda}\gamf\gamma_{\nu}T_{A,n, d g h\lambda}$ &  & 3 & $\epsilon^{ a b c}{\Bb_{ a}}^{ d}{u_{ b}}^{ e\mu}\nabla^{\nu}{{{f_{- c}}^{ f}}_{\nu}}^{\lambda}\gamf\gamma_{\lambda}T_{A,n, d e f\mu}$ & 80 & 324  \\
$\epsilon^{ a b c}{\Bb_{ a}}^{ d}{u_{ b}}^{ e\mu}{{u_{ c}}^{ f}}_{\mu}{u_{ e}}^{ g\nu}{u_{ g}}^{ h\lambda}\gamf\gamma_{\nu}T_{A,n, d f h\lambda}$ &  & 4 & $\epsilon^{ a b c}{\Bb_{ a}}^{ d}{u_{ b}}^{ e\mu}\nabla_{\mu}{f_{- d}}^{ f\nu\lambda}\gamf\gamma_{\nu}T_{A,n, c e f\lambda}$ & 81 & 325  \\
$\epsilon^{ a b c}{\Bb_{ a}}^{ d}{u_{ b}}^{ e\mu}{{u_{ c}}^{ f}}_{\mu}{u_{ e}}^{ g\nu}{u_{ g}}^{ h\lambda}\gamf\gamma_{\lambda}T_{A,n, d f h\nu}$ &  & 5 & $\epsilon^{ a b c}{\Bb_{ a}}^{ d}{u_{ b}}^{ e\mu}\nabla^{\nu}{{{f_{- d}}^{ f}}_{\mu}}^{\lambda}\gamf\gamma_{\nu}T_{A,n, c e f\lambda}$ & 82 & 326  \\
$\epsilon^{ a b c}{\Bb_{ a}}^{ d}{u_{ b}}^{ e\mu}{u_{ c}}^{ f\nu}{{u_{ d}}^{ g}}_{\mu}{u_{ e}}^{ h\lambda}\gamf\gamma_{\nu}T_{A,n, f g h\lambda}$ & 3 & 6 & $\epsilon^{ a b c}{\Bb_{ a}}^{ d}{u_{ b}}^{ e\mu}\nabla^{\nu}{{{f_{- d}}^{ f}}_{\nu}}^{\lambda}\gamf\gamma_{\mu}T_{A,n, c e f\lambda}$ & 83 & 327  \\
\hline\hline
\end{longtable}

\bibliography{references}
\end{document}